\documentclass[%
 reprint,
 amsmath,amssymb,
 aps,
]{revtex4-2}

\usepackage{graphicx} 
\usepackage{bm}
\usepackage[linktocpage=true,
  colorlinks=true, 
  pdfborder={0 0 0}, 
  linkcolor=blue,
  citecolor=red,
  filecolor=yellow,
  urlcolor=blue,
]{hyperref}

\begin{document}

\title{From Embeddings to Dyson Series: Transformer Mechanics as Non-Hermitian
Operator Theory}

\author{Po-Hao Chang}
\email{chang.pohao@gmail.com}

\affiliation{Quantum Science and Engineering Center, George Mason University, Fairfax,
VA 22030, USA}

\begin{abstract}
Transformer architectures are typically described in algorithmic and statistical terms, 
leaving their internal mechanics without a familiar structural language for researchers 
trained in physical theories. To bridge this gap,
we develop a complementary operator-theoretic
framework that recasts their mechanics in a language familiar to many-body
physics. Beginning from the token as a discrete index without intrinsic
geometry, we show that embedding corresponds to a basis transformation
into a continuous representation space. Once such a reference basis
is established, self-attention naturally assumes the role of a non-Hermitian
interaction operator, and network depth implements an ordered composition
of these interactions. Within this formulation, several empirical
properties of deep Transformers---including stability at large depth,
representational saturation, and the effectiveness of multi-head decomposition---find
natural structural interpretations as consequences of regulated operator
composition. Together, channel factorization and
normalization emerge as organizing structural logic rather than isolated
architectural choices. This perspective does not rely on post-hoc
analogy, but follows a constructive path where each parallel arises
from the preceding structural step. By recasting Transformer mechanics
in operator language, the framework lowers the conceptual barrier
between deep learning and many-body physics through shared mathematical
structure, making tools and intuitions from each domain more readily legible to the
other.
\end{abstract}
\maketitle

\section{Introduction: Two Different Universes}
  
Transformer architectures \cite{vaswani_attention_2017} have rapidly
become the dominant framework for large-scale representation learning.  
Their empirical success is well documented, and their mechanisms---self-attention,  
residual connections \cite{he_deep_2015}, normalization \cite{ba_layer_2016}, and multi-head decomposition \cite{vaswani_attention_2017}---are  
extensively analyzed within algorithmic and statistical paradigms.

Recent pioneering work has begun to bridge between physics and AI by applying statistical mechanics and random matrix theory to analyze the weight matrices of trained neural networks and characterize learning quality \cite{tishby_deep_2015,schoenholz_deep_2017,martin_implicit_2018,martin_setol_2025,pennington_resurrecting_2017}. 
However, these approaches primarily focus on the statistical or thermodynamic properties of the learned parameters themselves.
For researchers trained in operator-based physical theories, the dynamics of the sequence state, namely how the token vector propagates, interacts, and evolves through the Transformer’s depth, still lack a familiar structural language. The challenge is therefore not only one of statistical performance, but also of translating the architecture’s forward pass into a recognizable kinematic and dynamical framework.

Modern many-body physics and deep learning originate from fundamentally  
different intellectual starting points. 
In quantum mechanics, a system is bounded by its Hamiltonian $H$ constructed from explicit dynamical laws, which defines its allowable states. 
Machine-learning systems, in
contrast, are discovered through optimization: their structure emerges  
from data, with parameters adjusted to minimize a global loss over  
a statistical corpus whose generative rules are not explicitly defined.  

This difference in origin is not superficial. Physical operators inherit constraints, such as Hermiticity, unitarity, and conservation laws, which are non-negotiable consequences of physical laws. On the other hand, learned operators inherit no such constraints as properties like stability are enforced through deliberate architectural choices rather than first principles. The degrees of freedom available to each framework are therefore fundamentally different.
  
Despite this difference in origin, the resulting computational structures  
exhibit striking algebraic parallels. Both involve high-dimensional  
vector spaces, structured basis transformations, and repeated application  
of interaction-like operators that mix degrees of freedom across scales.
Furthermore, both frameworks mathematically involve resolving their high-dimensional states into probabilistic outcomes through either the measurement of a wave-function in quantum mechanics or a softmax projection over a vocabulary in language modeling.
From this perspective, the apparent conceptual gap between deep learning  
and many-body theory becomes less fundamental than it first appears.

In this work, we develop an operator-theoretic framework that recasts the internal mechanics of Transformers into the language of many-body physics. Starting from the token as a discrete, geometry-free index, we show that embedding acts as a basis transformation into a continuous representation space. Once this reference basis is established, self-attention naturally emerges as a non-Hermitian interaction operator, with network depth corresponding to an ordered composition of these interactions. 
 
Rather than drawing post-hoc analogies between machine learning and physics, we follow a constructive path where each parallel emerges from the mathematical structure of the architecture itself. Within this formulation, the empirical properties of deep Transformers, including stability at large depth, representational saturation, and multi-head effectiveness, acquire natural structural interpretations. Residual connections generate an exact algebraic expansion over ordered interaction paths, normalization regulates the magnitude of successive updates, and multi-head attention implements a channel factorization of the interaction operator. These elements thus emerge as organizing principles of operator composition rather than isolated heuristics. 

We assume familiarity with the basic Transformer components \cite{vaswani_attention_2017}, specifically self-attention and residual connections. We analyze their canonical Post-Norm forms here, as this provides the cleanest structural picture for the operator-theoretic framework, 
noting that modern architectural variations preserve these essential mechanisms despite differences in implementation. When invoking causal masking and the transfer-matrix-like generation mechanism, we draw on autoregressive variants whose directional interactions strictly enforce non-Hermitian operator structure. Bidirectional variants like BERT \cite{devlin_bert_2019} permit potentially symmetric attention, yielding operators closer to Hermitian, but are not the focus here

By expressing Transformer mechanics in operator language, this framework lowers the conceptual barrier between deep learning and many-body physics. The goal is not to force an analogy or claim exact mapping, but to expose a shared mathematical foundation—rendering the tools and intuitions of each field mutually legible. This perspective is intentionally incomplete. We emphasize only the most immediate structural parallels, and deeper examination will likely surface additional correspondences as well as further points of departure.

\section{The Kinematics: Embeddings as Basis Transformation}

In many-body physics, the starting point of any calculation is the
choice of basis. One typically begins with a discrete representation
namely, atomic orbitals \cite{ozakiVariationallyOptimizedAtomic2003,soler_siesta_2002},
plane waves \cite{kresseEfficientIterativeSchemes1996}, or site indices,
and transforms into a basis that captures the dominant low-energy
structure of the system. Diagonalization identifies stationary eigenstates,
selecting linear combinations that minimize the total energy within
the chosen variational subspace.

The Transformer begins from a mathematically analogous construction.
A token enters the model as a one-hot vector: a column vector of dimension $V$ with zeros everywhere except a single entry of one at the index corresponding to its vocabulary position, carrying no geometry beyond that discrete label, analogous to labeling a lattice site. 
The embedding matrix $W_{E}$ performs
a learned linear transformation on this index space, projecting the
discrete label $e_{i}$ into a continuous latent vector $x_i^{(0)}$of dimension
$d_{\mathrm{model}}$ \cite{mikolov_efficient_2013,devlin_bert_2019}:
\begin{equation}
\underbrace{x_i^{(0)}}_{\mathbb{R}^{d_{\text{model}}}}
=
\underbrace{W_E}_{d_{\text{model}} \times V}
\underbrace{e_i}_{\mathbb{R}^{V}}\textbf{} 
\end{equation}


Because $d_{\mathrm{model}}\ll V$,  where $V$ is the dimension of the vocabulary space, this projection assigns each discrete vocabulary index a vector coordinate in a lower dimensional representation space
---a downfolding in which only the dominant relational structure is retained
and distinct discrete identities are mapped into a shared continuous
geometry. 

The procedure is loosely analogous to a variational reduction \cite{book},
where the massive discrete space is projected out and only the relevant
low-lying active subspace is kept.

The objectives differ: electronic structure minimizes the Hamiltonian
expectation value, while the embedding matrix is learned through predictive
loss over a training corpus. The structural role is nonetheless comparable
as both identify optimal linear combinations within a constrained
basis that best represent the underlying structure of the system.
The resulting vectors $x^{(0)}$ therefore serve as stationary reference
states for the vocabulary, defining the kinematic manifold prior to
contextual interaction. Subsequent layers then introduce effective interactions
that mix these reference states, generating context-dependent representations
analogous to correlated many-body states built upon an unperturbed
eigenbasis.

\section{The Dynamics: Perturbation as a Mental Picture for Attention}

\begin{figure}
\includegraphics[scale=0.29]{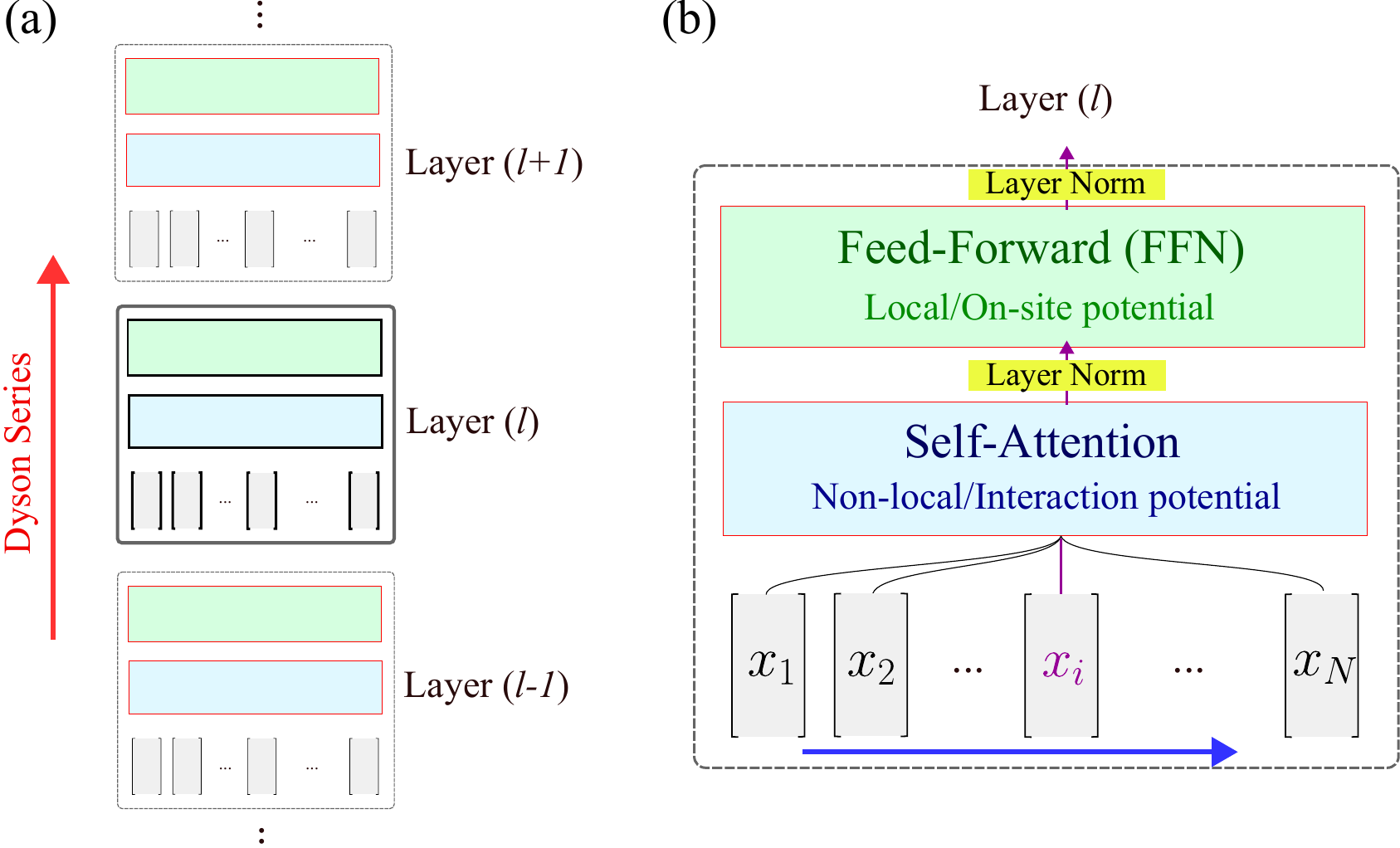}\caption{\label{fig:transformer} 
Schematic representation of the Transformer
architecture. (a) Transformer layers are depicted as discrete evolution
steps along the vertical axis (red), representing ordered layer depth.
(b) Detailed view of an individual $l$-th layer: the Self-attention
block introduces non-local, off-diagonal coupling across the lattice
sites, while the Feed-Forward Network (FFN) acts as a local, on-site
operator. The forward pass through the network corresponds to the
successive application of these operators, analogous to an evolution
process.}
\end{figure}

To understand how the Transformer evolves the state of a sequence \cite{vaswani_attention_2017}, 
we must first define its core computational block. The architecture operates through the 
repeated application of a modular structure: as shown in Fig. \ref{fig:transformer} (a), it consists of a stack of $L$ structurally identical layers, each with distinct learned weights, applied sequentially to (b) a system of $N$ degrees of freedom (token states) within a layer.

\subsection{The Transformer modular unit: Non-Local vs. Local Operators}

Each layer in the network, as shown in Fig. \ref{fig:transformer}
(b), can be viewed as a discrete evolution step for a ``multi-particle
system of $N$ tokens''. Every layer is composed of two primary computational
components that serve distinct physical roles:
\begin{enumerate}
\item \textbf{The Self-Attention Block (Non-Local Coupling):} This component depicted as blue rectangle in Fig. \ref{fig:transformer}(b),
acts as the interaction engine. It allows every token (state) $x_{i}$
in the sequence to communicate with other tokens $x_{j}$. In the
language of operators, this is an off-diagonal process that couples
different tokens across the sequence, enabling the exchange of semantic
information.
\item \textbf{The Feed-Forward Network (Local Potential): }Following the
interaction, each token state passes through a position-wise Feed-Forward (FFN) shown as green rectangle in Fig. \ref{fig:transformer}(b). This
is a purely local operator. It does not couple different tokens; instead,
it acts like a learned on-site potential that encodes integrated statistical structure from training \cite{geva_transformer_2021,meng_locating_2023}, analogous to how pseudopotentials fold core electron complexity into an effective valence-space operator \cite{vanderbiltSoftSelfconsistentPseudopotentials1990}.
\end{enumerate}
The entire "forward pass" of the model
is simply the iterative application of these two operators defined
in Fig. \ref{fig:transformer}(b) and (a), respectively,
supported by intermediate Layer Normalization steps (yellow blocks) 
that stabilize the state vector. This alternating
sequence of non-local mixing and local refinement is what allows the
model to build complex, context-dependent representations from initially
independent basis states (i.e. static embedding).

\subsection{The Heart of Attention: Q, K, and V as Coupling Components}

The interaction within the attention block are defined by three learned
linear projections \cite{vaswani_attention_2017}:
\[
Q=W_{Q}x,\;K=W_{K}x,\;V=W_{V}x
\]
These projections establish a coupling protocol between token sites. 

The Query ($Q_{i}$) acts as a "receptor"
for state $i$, determining what kind of surrounding states it needs
to interact with. Conversely, the Key ($K_{j}$) acts as the "signature"
of state $j$, defining exactly how it appears to those probing receptors.

For a pair of states $i$ and $j$ the interaction amplitude is determined
by the overlap
\[
w_{ij}\propto Q_{i}^{T}K_{j}
\]
which serves as an effective coupling coefficient. This quantity controls
how strongly the state component associated with site $j$ contributes
to the update of site $i$.

Finally, the value projection $V_{j}$ specifies which vector components to
be transferred under this coupling. The full update maybe schematically
expressed as 
\[
x_{i}^{\text{new}}=x_{i}+\sum_{j}w_{ij}V_{j}
\]

In operator terms, $W_{Q}$ and $W_{K}$ define the interaction kernel
(coupling strengths) while $W_{V}$ defines what state content
to be mixed. Together they implement a structured, non-Hermitian interaction
that couples otherwise independent sites.

\subsection{Self-Attention as an Effective Interaction}

Self-attention mechanism introduces coupling between tokens through an effective operator in the layer $l$
$$V_{\text{eff}}^{(l)}\propto W_{Q}^{T}W_{K}.$$
Unlike conventional quantum Hamiltonians, $V_{\text{eff}}^{(l)}$ is generically non-Hermitian. In standard quantum mechanics, Hermiticity ensures that the coupling between states $i$ and $j$ satisfies $\langle\psi_{j}|\hat{V}|\psi_{i}\rangle=\langle\psi_{i}|\hat{V}|\psi_{j}\rangle^{*}$. 
This symmetric coupling, combined with an evolution operator containing an explicit imaginary factor ($I - i\hat{H}dt$), guarantees unitary time evolution, preserves probability amplitude, and drives oscillatory dynamics. 

Language, however, is intrinsically directional, a property further enforced in autoregressive models \cite{brown_language_2020,radford2018improving} by the causal attention mask. This mask imposes left-to-right dependencies across the 1D sequence token lattice (blue arrow, Fig. \ref{fig:transformer} (b)). While all $N$ states evolve simultaneously through the network depth, state $i$ can only be perturbed by spatially preceding states $j \le i$. This spatial constraint renders the $N \times N$ effective interaction matrix strictly lower-triangular, permanently breaking reversibility. 

Furthermore, the Transformer's residual mechanism strictly lacks the imaginary unit, acting instead as a purely real generator. Consequently, the layer update is non-unitary. 
Rather than exhibiting oscillatory dynamics, representations are iteratively reshaped through a non-unitary, irreversible forward pass.

This purely real, non-unitary evolution mathematically invites comparison to imaginary-time evolution, where the transformation $\tau = it$ absorbs the phase to yield amplitude relaxation. However, the analogy is only precise in architectures with weight tying, such as Universal Transformers \cite{dehghani_universal_2019} or Deep Equilibrium Models \cite{bai_deep_2019}, where ${V}^{(l)}_\text{eff}={V}_\text{eff}$. Imaginary-time evolution presumes a fixed operator $\hat{H}$, whereas in standard Transformers, each layer carries distinct learned parameters, so the effective interaction ${V}^{(l)}_\text{eff}$ varies with depth. 

For standard Transformers with distinct layer weights, the evolution
should instead be viewed as a non-autonomous operator flow. The imaginary-time
picture remains a useful heuristic for understanding why representations
saturate across depth and why deeper networks do not generically diverge,
but it does not imply relaxation toward the ground state of a single
fixed Hamiltonian.

%
%
\subsection{Multi-Head Attention as Channel Factorization}
\begin{figure}
\includegraphics[scale=0.34]{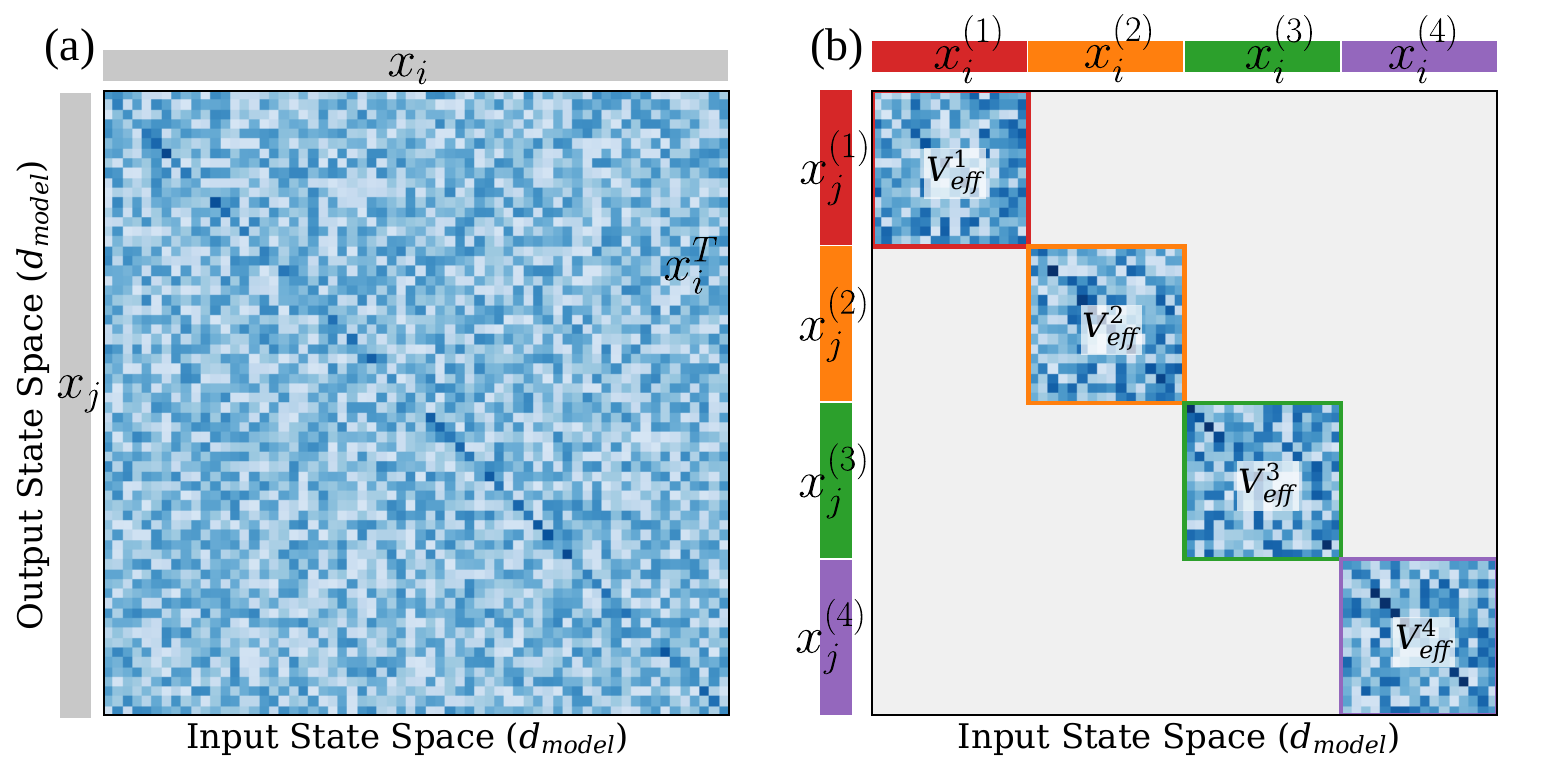}
\caption{
\label{fig:mha}
Schematic view of multi-head attention as operator channel factorization.
(a) A dense effective interaction $V_{\text{eff}}$ operating on the full representation space $d_{model}$ to map input states $x_j$ to updated states $x_i$. (b) In multi-head attention, the interaction is block-diagonalized into $h$ independent channels ($h=4$ in this example). The state vectors themselves are partitioned into corresponding reduced subspaces ($x_j^{(h)}$ and $x_i^{(h)}$). Each block $V_{\text{eff}}^{(h)}$ operates exclusively within its own sub-vector.
}
\end{figure}

In practical Transformer architectures, the interaction operator is not implemented as a single coupling channel as shown in Fig. \ref{fig:mha}(a). Instead, multi-head attention decomposes the interaction into several ($h$ in this case) parallel components acting on reduced representation subspaces as depicted in Fig. \ref{fig:mha}(b) \cite{vaswani_attention_2017}. Each head performs an independent attention operation with its own projection matrices, producing an update that is later recombined through linear projection.

From an operator perspective, this architecture factorizes the effective interaction into multiple channels,
%
$$V_{\text{eff}} = \begin{pmatrix} V_{\text{eff}}^{(1)} & 0 & \cdots & 0 \\ 0 & V_{\text{eff}}^{(2)} & \cdots & 0 \\ \vdots & \vdots & \ddots & \vdots \\ 0 & 0 & \cdots & V_{\text{eff}}^{(h)} \end{pmatrix}$$
where each component $V_{\text{eff}}^{(k)}$ ($k=1,2,$ ... $h$) acts within a restricted representation subspace before their contributions are aggregated.

This structure resembles operator decompositions commonly used in many-body physics, where distinct interaction channels probe different sectors of the system. Examples include separations between spin and spatial operators or decompositions into angular-momentum channels in partial-wave expansions. 
Although Transformer heads operate within a shared representation space rather than independent Hilbert spaces, their architectural role is similar: structured operators extract complementary information from different projections of the state before recombination.

In this sense, multi-head attention implements a controlled factorization of the interaction operator, allowing complex coupling patterns to be constructed from several lower-dimensional channels. Empirical and theoretical work has shown that self-attention admits low-rank approximations \cite{wang_linformer_2020}, supporting the interpretation of multi-head structure as a factorized representation of the underlying interaction.

\subsection{The Residual Stream as First-Order Perturbation }

\begin{figure}
\includegraphics[scale=0.4]{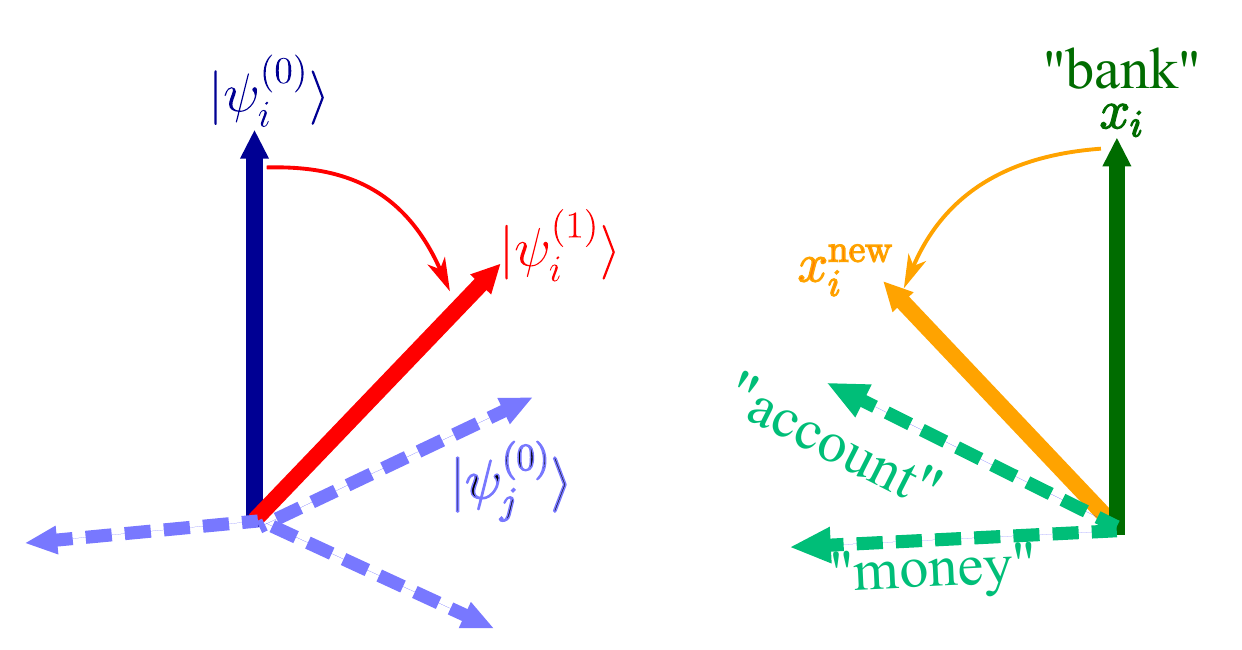}
\caption{
\label{fig:perturbation}
Geometric mapping between first-order perturbation
theory and contextual token mixing. (Left) An unperturbed quantum
reference state $|\psi_{i}^{(0)}\rangle$ is dressed by surrounding
basis states $|\psi_{j}^{(0)}\rangle$ to form the perturbed state
$|\psi_{i}^{(1)}\rangle$. (Right) Analogously, an ambiguous static
token embedding $x_{i}$ (e.g., "bank" )
is contextually resolved by mixing in representations from preceding
tokens (e.g., "account", "money" )
to produce the updated state $x_{i}^{\mathrm{new}}$
}
\end{figure}

Stripped of engineering stabilizers, the core residual update 
for an autoregressive token incorporates a causal attention mask that
acts as a physical Heaviside step function $\Theta(i-j)$, restricting
the inter sum strictly to preceding states: 
\[
x_{i}^{\text{new}}=x_{i}+\sum_{j=1}^{i}(x_{i}^{T}V_{\text{eff}}x_{j})x_{j}.
\]
This non-reciprocal spatial coupling ensures that each token state is "dressed"
exclusively by the interactions with its leftward lattice neighbors,
inherently accommodating the sequential addition of new basis states
without altering the computed history of prior tokens. This mechanism
resolves contextual ambiguity. A token such as " bank",
initially represented as a superposition of semantic possibilities,
is driven toward the subspace consistent with its environment through
effective interaction with neighboring states as schematically depicted
in Fig. \ref{fig:perturbation}. 

This expression mirrors the structure of first-order interaction mixing 
in quantum mechanics,
\[
\psi_{i}^{(1)}=\psi_{i}^{(0)}+\sum_{j=1}^{N}\langle\psi_{i}^{(0)}|\hat{V}_{\text{int}}|\psi_{j}^{(0)}\rangle\psi_{j}^{(0)}
\]
where $\hat{V}_{\text{int}}$ is the interaction operator that induces mixing between the 
unperturbed states. 
In this sense, the residual update corresponds to the first-order interaction step in an operator expansion. Despite sharing a similar algebraic form with the stationary-state corrections of Rayleigh–Schr\"odinger perturbation theory, it is mathematically more analogous to the leading term of a Dyson-type evolution.

The residual term $x_{i}$ plays the role of the unperturbed reference
state $|\psi_{j}^{(0)}\rangle$. The surrounding tokens $x_{j}$ form
the basis states mixed into the target state, weighted by interaction
amplitudes determined by the attention scores.

\subsection{From Theory to Engineering Practice}

In practice, the interaction amplitudes are not taken directly from the raw overlaps $Q_i^TK_j$. Instead, the Transformer defines a normalized interaction kernel that regulates the magnitude of the perturbative mixing. The overlaps are first scaled by $1/\sqrt{d_k}$, where $d_k$ is the dimensionality of the key and query vectors. This factor ensures that the variance of the interaction amplitudes remains approximately constant as the representation dimension grows. 

The resulting matrix is then passed through a softmax function along the interaction index $j$, producing a set of normalized coupling weights that sum to unity for each receiving site $i$. Mathematically this operation resembles a Boltzmann distribution, where the scaled overlaps act as interaction “energies” and the factor $\sqrt{d_k}$ sets an effective temperature controlling how sharply the interaction selects among competing states. In this way the raw interaction strengths are converted into a bounded set of mixing coefficients that regulate the perturbative update.

The value projection $W_V$ further separates the static identity of a token from the information being transferred through the interaction, enabling controlled state mixing across sites.

With these definitions, the practical self-attention update takes the familiar form

\begin{equation}
x_{i}^{\text{new}} = x_{i} + \sum_{j=1}^{N}\text{softmax}\left(\frac{Q^{T}K}{\sqrt{d_{k}}}\right)_{ij}V_{j}
\label{eq:attention}
\end{equation}
These normalization steps ensure that the perturbative mixing remains bounded even for large sequence lengths and high-dimensional representations, allowing the interaction operator to be evaluated stably at scale. 

Crucially, this practical attention mechanism departs from the strictly linear operators familiar in conventional quantum mechanics, as the softmax function and $1/\sqrt{d_k}$ scaling introduce a highly nonlinear dependence of the interaction weights on the state.

However, it is worth recalling the fundamentally distinct nature between physical and learned operators that governs all such departures. Unlike a physical system constrained by immutable laws, in a software model, an engineer has the mathematical freedom to impose stability directly. Consequently, the practical stabilizers introduced here are best understood as engineered regulators whose structural function serves as the relevant point of comparison to physical constraints.

Despite these modifications, the conceptual paradigm remains intact: self-attention functions as a non-Hermitian operator enabling state mixing, while its normalization mechanisms serve as engineering stabilizers. Specifically, $1/\sqrt{d_k}$ scaling maintains $O(1)$ interaction variance \cite{vaswani_attention_2017}, 
the softmax ensures convex combinations of value vectors, and Layer Normalization, depicted as the yellow block shown in Fig. \ref{fig:transformer} (b), rescales activations to a controlled variance \cite{ba_layer_2016,xiong_layer_2020}. Effectively, the bare residual update is structurally modified into a regulated sublayer operation. If we denote the entire non-linear attention summation in Eq. \ref{eq:attention} as an interaction functional $\mathcal{A}(x_i) \equiv \sum_{j=1}^{N}\text{softmax}\left(Q^{T}K / \sqrt{d_{k}}\right)_{ij}V_{j}$, the standard attention sublayer update takes the form $x_{i}^{\prime} = \text{LayerNorm}(x_{i} + \mathcal{A}(x_{i}))$. 
When subsequently combined with the local FFN potential, the complete forward pass through layer $l$ can be mathematically lumped into a single composite effective state-dependent operator $\hat{V}^{(l)}$, yielding the stable recurrence 
\begin{equation}
x_i^{(l)} = x_i^{(l-1)} + \hat{V}^{(l)}x_i^{(l-1)}
\label{eq:recurrence}
\end{equation}
Residual connections implement the identity shortcut \cite{he_deep_2015}, with normalization further stabilizing the operator composition throughout training \cite{de_batch_2020}, collectively ensuring a well-regulated forward pass.

\section{The Evolution of Depth: Time-Ordered Dyson Series}

If a single-layer update mirrors a first-order perturbation, then the stacking of distinct operators across depth \cite{veit_residual_2016, greff_highway_2017} organizes the forward pass as an ordered composition equivalent to evaluating successive, higher-order many-body corrections.

Consider a deep Transformer where each layer $l$ possesses unique
projection matrices, meaning the effective potential $\hat{V}_{\text{eff}}^{(l)}$
is layer-dependent. To a physicist, the depth of the neural network
acts as a discrete time axis, where the sequence state propagates
through a time-varying interaction Hamiltonian.

Let the static token embedding be the initial, unperturbed state $x^{(0)}$.
The first residual layer computes a standard first-order update: 
\begin{equation}
x^{(1)}=x^{(0)}+\hat{V}^{(1)}x^{(0)}=(I+\hat{V}^{(1)})x^{(0)}.\label{eq:residual_connect}
\end{equation}
The profound structural recurrence emerges in the second layer. The
network applies the second potential $\hat{V}^{(2)}$ not to the base
state, but to the already perturbed state $x^{(1)}$: 
\begin{equation}
x^{(2)}=x^{(1)}+\hat{V}^{(2)}x^{(1)}=(I+\hat{V}^{(2)})(I+\hat{V}^{(1)})x^{(0)}
\end{equation}

Expanding this algebraic product reveals the exact generation of higher-order
interaction terms: 
\begin{equation}
x^{(2)}=\left[I+\hat{V}^{(1)}+\hat{V}^{(2)}+\hat{V}^{(2)}\hat{V}^{(1)}\right]x^{(0)}.
\end{equation}
The term $\hat{V}^{(2)}\hat{V}^{(1)}$ represents a mathematically
second-order correction, where the state is perturbed by Layer 1 and
subsequently perturbed again by Layer 2. By induction, the full forward
pass through $L$ layers mathematically evaluates the discrete product
operator: 
\begin{equation}
x^{(L)}=\prod_{l=1}^{L}(I+\hat{V}^{(l)})x^{(0)}
\end{equation}

When expanded, this product generates a sum over all possible ordered
compositions of the interaction operators \cite{veit_residual_2016}:

\begin{equation}
\begin{aligned}x^{(L)}=\bigg[I & +\sum_{l=1}^{L}\hat{V}^{(l)}+\sum_{l_{2}>l_{1}}^{L}\hat{V}^{(l_{2})}\hat{V}^{(l_{1})}\\
 & +\sum_{l_{3}>l_{2}>l_{1}}^{L}\hat{V}^{(l_{3})}\hat{V}^{(l_{2})}\hat{V}^{(l_{1})}+\dots\bigg]x^{(0)}
\end{aligned}
\label{eq:nn_series}
\end{equation}

In quantum field theory, this mirrors the discrete formulation of
the time-ordered Dyson series \cite{Sakurai1993Modern}:

\[
\begin{aligned}U(t,t_{0}) & =\mathcal{T}\exp\left(-i\int_{t_{0}}^{t}\hat{V}(t')dt'\right)\\
 & =I-i\int_{t_{0}}^{t}\hat{V}(t')+(-i)^{2}\iint_{t_{2}>t_{1}}\hat{V}(t_{2})\hat{V}(t_{1})\\
 & +(-i)^{3}\iint_{t_{3}>t_{2}>t_{1}}\hat{V}(t_{3})\hat{V}(t_{2})\hat{V}(t_{1})+\dots
\end{aligned}
\]

For brevity, we adopt a shorthand notation where the differential time measures $dt_n$ and the shared integration limits $[t_0, t]$ are omitted from the expansion.

%
%
While this compositional ordering mirrors the structure of the time-ordered Dyson series, it is critical to recognize that the operators $\hat{V}^{(l)}$ are nonlinear functionals, driven by the softmax and feed-forward activation functions. Consequently, the expanded cross-terms (e.g., $\hat{V}^{(2)}\hat{V}^{(1)}$) represent a composition of nonlinear mappings rather than standard linear matrix products. This places a formal boundary on the analogy: these composite terms cannot be directly analyzed via standard linear spectral decomposition without employing careful linear approximations.

It is interesting to note that state-dependent operators of this kind have a natural counterpart in time-dependent density functional theory (TDDFT) \cite{gross1984}, where the effective Hamiltonian $\hat{H}[\rho(t)]$ similarly depends on the evolving state at each time step. The standard Transformer in fact shares both features, state-dependent operators and layer-varying parameters, making TDDFT a potentially richer analogy than the static Dyson series. However, as the goal of this work is to establish a clear structural foundation, we retain the simpler time-ordered operator picture and leave the TDDFT correspondence to future investigation.

Through this iterative structure, network depth generates higher-order, "time-ordered" many-body correlations. Linguistically, the architecture scales from coupling isolated words to interacting phrases and complex clauses, iteratively building perturbative corrections until the state converges.

\subsection{Layer Norm as Wavefunction Renormalization}

While high-order perturbation expansions in quantum systems frequently
diverge, 
The Transformer architecture helps maintain stability through an iterative “re-gauging” mechanism known as Layer Normalization \cite{ba_layer_2016}. Physically, this plays a role analogous to a renormalization-like rescaling applied at each order of the expansion. Rather than allowing successive operator compositions to drive the state vector’s norm into uncontrolled regimes, LayerNorm continuously rescales the dressed state, keeping the depth expansion within a bounded region of representation space.

It effectively acts as a stability regulator for the operator composition, empirically observed in by models like GPT-3 \cite{brown_language_2020}, where architectures with up to 96 layers maintain coherent representations across depth.
The exact placement of this renormalization step and its precise form vary across modern architectures, but the structural role as a stability regulator of the operator composition remains unchanged.

\section{CAUSALITY AND MEASUREMENT: THE SPATIAL ARROW AND UNEMBEDDING}

Autoregressive generation functions as a transfer-matrix-like procedure on a 1D directed token lattice. At each step, the sequence boundary advances by one site: the "dressed" edge state is measured, a discrete outcome is then appended, and the boundary state is updated. This process bridges the two primary axes of the architecture depicted in Fig. \ref{fig:transformer}: the computational depth of the forward pass and the spatial growth of the sequence.

The forward pass serves to prepare the boundary state $|\Psi_{N}\rangle$ for measurement. By the final layer $L$, the token at position $N$ has evolved into a fully "dressed" state, encoding the contextual influence of all preceding tokens through non-local interactions and local renormalizations. 
Because of this causal structure, $|\Psi_{N}\rangle$ acts as a sufficient representation for the next-token prediction and no global measurement of the full sequence is required.

The unembedding matrix $W_{U}$ operates as the measurement operator, projecting the boundary state back onto the discrete vocabulary basis $\langle v_{k}|$. This projection yields the transition amplitudes:$$\text{Logits}_{k} = \langle v_{k} | \Psi_{N} \rangle$$The structural parallel to quantum measurement is mathematically precise: both involve projecting a high-dimensional state onto a discrete basis and normalizing the resulting magnitudes into a probability distribution.

However, this mapping is physically distinct. While quantum probabilities emerge from  the geometry of Hilbert space via squared wavefunction amplitudes, 
This framework emphasizes the shared algebraic sequence, projection followed by normalization, rather than suggesting a common physical origin.

\section{CONCLUSION AND OUTLOOK}

Looking at Transformer mechanics through the lens of basis construction,
effective interactions, and ordered state mixing provides more than
a metaphor. It offers an effective mental model. For researchers trained
in computational physics, these architectures need not appear as opaque
statistical engines; they can be understood as structured operator
systems evolving within a variationally optimized subspace.

The operator-theoretic bridge proposed here, however, rests on a structural
parallel rather than an exact identity mapping. 
While physical operators follow from first principles, such as Hermiticity and unitarity, 
Transformer weights are empirical outcomes of gradient descent and nonlinear in nature. 
Despite these differing origins, both disciplines confront similar structural bottleneck:
the stabilization of large ordered products of non-commuting, non-Hermitian
operators against divergence or rank collapse --- an instability
independent of whether the constituent operators arise from a Hamiltonian
or a learned weight matrix.

The transferability of techniques depends on identifying which operator properties govern this composition. 
These are not Hermiticity or unitarity, but the non-Hermitian, dense, and potentially defective character of individual layers. 
The matrices underlying these transformations share structural features with the classes of operators studied in non-Hermitian random matrix theory \cite{saada_mind_2025,staats_small_2025} and pseudospectral analysis \cite{trefethen_spectra_2005}. While the present work does not establish a direct mapping, these frameworks may provide natural tools for diagnosing stability and sensitivity in deep operator compositions.

While the constituent layer operations are nonlinear, the depth expansion in Eq. \ref{eq:nn_series} generates a summation over state vectors within the linear representation space. For architectures that seek fixed points, such as Deep Equilibrium Models \cite{bai_deep_2019}, fixed-point acceleration methods such as Anderson mixing, the numerical analog of DIIS \cite{pulay_convergence_1980}, are already employed in practice. For standard Transformers with depth-varying operators, sequence extrapolation methods operating directly on the state trajectory, such as vector Pad\'e and Borel \cite{borel_memoire_1899} resummation, may offer complementary tools for extracting stable representations from this discrete series, though their applicability to nonlinear compositions requires further investigation.

The transfer of knowledge between these fields is inherently reciprocal.
From a physics perspective, the empirical stability of large-scale
Transformers is remarkable. 
They maintain coherent state vectors across many sequential, non-Hermitian layer operations without catastrophic divergence in practice, despite being shaped by optimization across billions of scalar parameters. Deep learning achieves this not through physical conservation laws, but through structural interventions like residual connections and layer normalization that act as algorithmic regulators for strongly out-of-equilibrium dynamics. 

As discussed earlier, deep learning has the mathematical freedom to rescale state vectors through architectural choices to enforce stability that is prohibited by the conservation laws governing physical quantum systems. 
However,  this divergence in constraints is what makes cross-disciplinary exchange valuable. Translating AI’s empirically successful stabilization mechanisms into operator language may offer physicists new perspectives on controlling complex non-Hermitian dynamics. Conversely, the deep learning community can draw upon the analytical frameworks of many-body physics to better characterize the stability and structure of these architectures.

By identifying operator product control as the shared numerical bottleneck,
we move beyond analogy toward systematic translation. The value of
this perspective lies not in providing immediate solutions, but in
establishing that physicists and machine learning researchers confronting similar mathematical 
and structural bottlenecks,  and can therefore begin working on them together.

\bibliographystyle{apsrev4-2}
\bibliography{refs.bib}

\end{document}